# Landscape of atomic nuclear shapes

Chang-Bum Moon*

*Hoseo University, Chung-Nam 336-795, Korea*

June 13, 2016

We exhibit a wide variety of the nuclear shape phases over the nuclear chart along with a shell model scheme. Various nuclear shapes are demonstrated within the framework of proton-neutron spin-orbital interactions; ferro-deformed, sub-ferro-deformed, and spherical shapes. The spherical shape is classified into the three magic-number categories in view of a large shell gap mechanism; double-magic nuclei I, double magic nuclei II, and double magic nuclei III. We discuss nuclear shape coexistence in the space Z = 76 to 84 as providing a new way to understanding the dynamical shape phases.

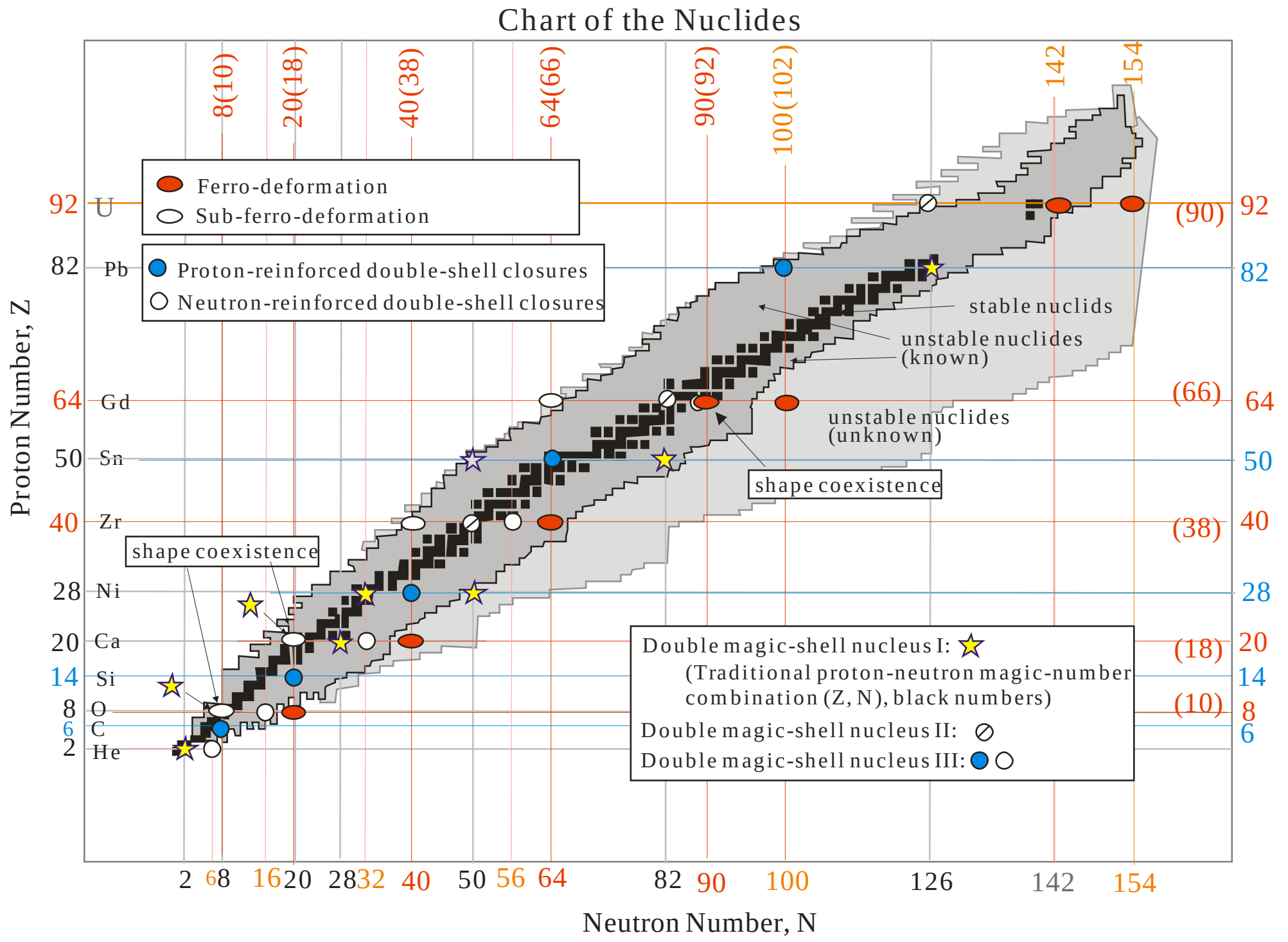

*cbmoon@hoseo.edu

## 1. Introduction.

Henri Poincaré tells us: “The scientist studies nature because he takes pleasure in it, and he takes pleasure in it because it is beautiful. If nature were not beautiful, it would not be worth knowing, and life would not be worth living. Of course I am not speaking of that beauty which strikes the senses, of the beauty of qualities and appearances. I am far from despising this, but it has nothing to do with science. What I mean is that more intimate beauty which comes from the harmonious order of its parts, and which a pure intelligence can grasp.” [1]. In natural science, two types of order play a essential role; patterning and organization. Patterning is relevant in science because empirical control is achieved through experiments whose possibility depends on the very existence of patterns in the natural world. Science is centered on the search for spatiotemporal patterns. Although nature does not consist only of patterns, it is organized around configurations (spatial patterns) and rhythms (temporal patterns) [2]. The atomic nucleus is one of the organized quantum systems, emerging outstanding patterns; both configurations and rhythms.

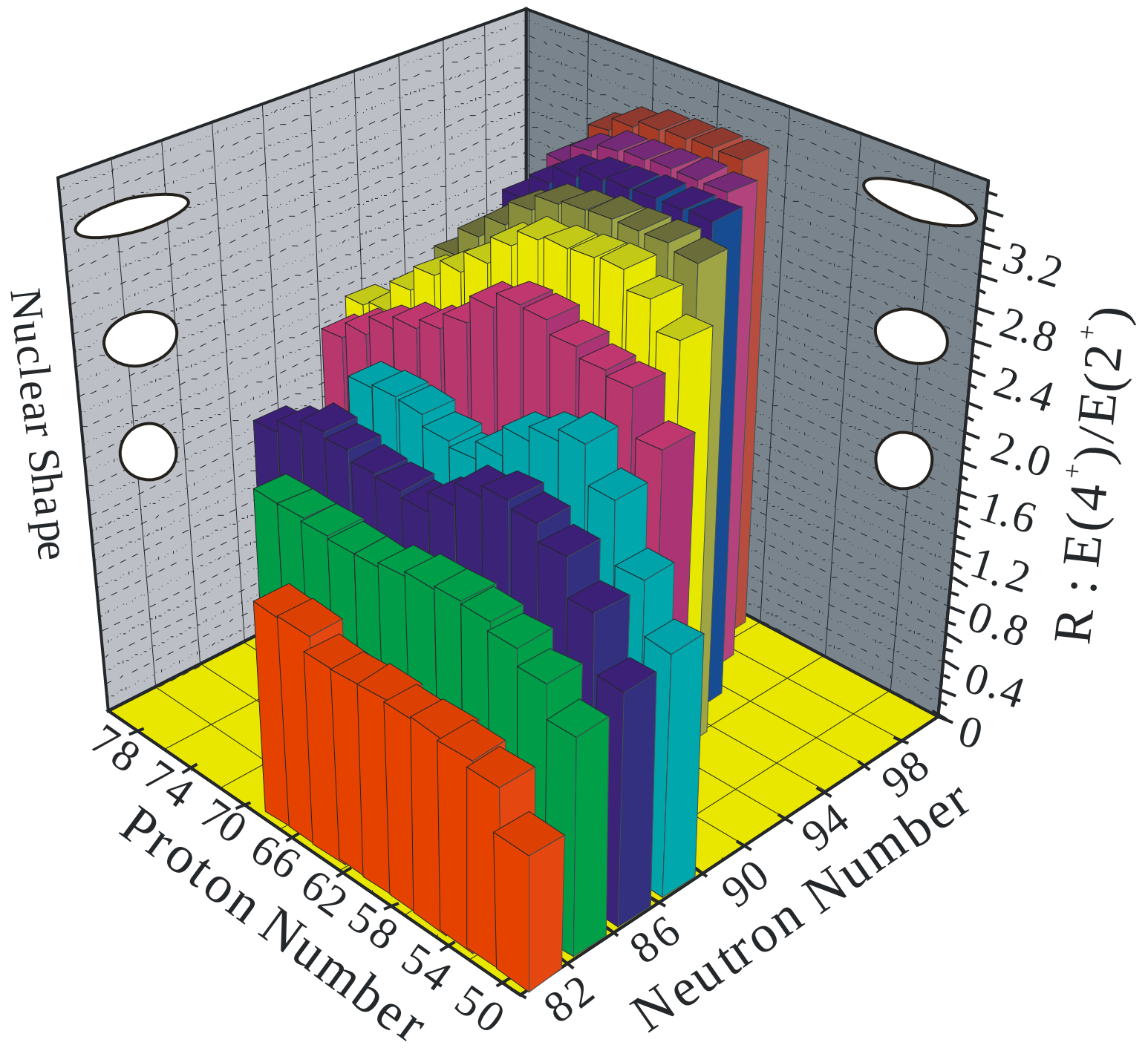

Fig. 1. Illustration of the nuclear structure patterns indicating shape changes in the order parameter R according to the constituent proton and neutron numbers. The R parameter, as an indicator of nuclear shapes, corresponds to the energy ratio of the first excited $2^+$ and the $4^+$ states in even-even nuclei. We can see that revealed is patterning progressively from no peak through two peaks to one peak along the neutron number. Nuclear shapes are commonly classified as a spherical, a spheroidal, and an ellipsoidal symmetry.

Figure 1 demonstrates the emerging patterns of nuclear shapes in terms of a nuclear deformation parameter, indicating spherical and non-spherical (deformed) according to the number of organized nucleons; protons, Z, and neutrons, N. The shell structure of atomic nuclei is a good example of natural patterns and one of the cornerstones for a comprehensive understanding of the quantum world.

In this work, we exhibit the nuclear shape patterns as outlining nuclear structures over the nuclear chart. On the basis of proton-neutron spin-orbital interaction schemes, various nuclear shapes are demonstrated orderly and systematically; ferro-deformed, sub-ferro-deformed, and spherical shapes. We classify the double-magic number nuclei in order to make sure of spherical closed-shell magic number definition. Finally, we discuss dynamical nuclear shape coexistence in the space Z = 76 to 84.

## 2. Classification of nuclear shapes.

The nuclear magic number is a representative phenomenon in the quantum world. These values correspond to filled proton or neutron shells, which looks like the periodicity of chemical properties of the elements. By introducing a strong spin-orbital coupling term into the single particle potential, the magic numbers could be reproduced [3, 4]. It means that the spin-orbital interaction in a nucleus plays a central role in controlling nuclear structures. Figure 1 illustrates the nuclear shell model scheme that produces the magic numbers; 2, 8, 20, 28, 50, 82, and 126. However, as being far off the stability regime, the magic numbers cannot fit any longer to a spherical shell gap but, rather yield non-spherical structure, indicating a spherical shell-gap breaking. With concepts of the proton-neutron spin-orbital interactions in previous works [5-8], we were able to explain the occurrence of such a (large) deformation by introducing the pseudo-shell configuration. The very deformation was classified, as being named a ferro-deformation, orderly on the (Z, N) coordinates; (8, 20), (20, 40), (40, 64), (64, 102). Notice that the numbers N, Z = 18, 38, 66 are also included in these categories. We demonstrate distributions of the ferro-deformation on the shell model diagram and on the chart of the nuclides in Fig. 2(a) and 2(b), respectively. For the case of (64, 102), the ferro-deformation is broadly distributed over (64, 92) to (74, 110) as being centered at (66, 104) [5]. The reason is mainly from a huge pseudo-shell formation combined by all the sub-shell orbitals, $J_{dghs}(31/2)$ in the proton 50 to 82 and $J_{hfpi}(43/2)$ in the neutron 82 to 126, respectively. See Fig. 4 in [5] and Fig. 3 in the present work.

In addition to the ferro-deformation, we introduce a sub-ferro-deformation built at the same number proton-neutron combination such that; (20 or 18, 20 or 18), (40 or 38, 40 or 38), (64 or 66, 64 or 66). Here the coordinates (40 or 38, 40 or 38) provide four possible configurations, such as (40, 40), (40, 38), (38, 38), (38, 40). This region, for the Sr and Zr nuclei, is well known to be observed the very deformed structure [9, 10]. If we look at the data of the (20, 20) system at the national nuclear data center (NNDC) [11], $^{40}Ca$, we might find a developed collective structure built on the second $0^+$ state as though it exhibits a large spherical-shell gap structure on the ground state.

When focusing on the opposed side of shell levels with respect to the ferro-deformation, we find a shell-gap reinforcement from the result of proton-neutron spin-orbital interactions. For instance, look at the Z = 14 shell gap and the N = 16 gap. Taking the $\nu d_{3/2}$-$\pi d_{5/2}$ interaction into account, it is expected that both of two shell gaps become more widen, leading to a double shell closure at Z = 14, N = 16. But it is not the case. Instead, the double-shell like closures appear to be at (14, 20) and (8, 16), respectively. This indicates that a spherical double-shell closure occurs at an optimized condition in order that the magic number shell gap Z = 8 or N = 20 might be involved. This is the case that is the combination of (14, 20) or (8, 16). Sometimes, this double-shell gap enlargement is called a new magic number or a magic nucleus. We define these two types of double-shell closures; on one hand, the Z = 14, N = 20 is a proton reinforced shell gap and on the other, the Z = 8, N = 16 is a neutron reinforced shell gap. We classify both double-shell gaps as the double-magic nuclei III.

If we regard double-shell gaps as a general phenomenon due to the spin-orbital interactions between proton-proton, neutron-neutron, and proton-neutron, it would not be necessary to use the specific words for describing shell structures such as a *new*, a *disappearance*, a *quenching*, an *exotic* etc. We have to notice that the spin-orbital interaction operates for yielding a large deformation as well, even at the magic numbers. It is in a common sense that we might as well categorize the emerging nucleon shell patterns on the chart of the nuclides like the periodic table of the elements. Figure 2 summarize our categorization for nuclear shape descriptions on the shell model-level scheme and on the chart of the nuclides.

If we turn our attention to the magic numbers 28, 50, and 82, situations are quite different and even subtle comparing to those at 8, 20, and 40, which correspond to the shell gap numbers built on the harmonic oscillator potential without spin-orbital interactions. In other words, the presence of the spin-orbital interactions is the Raison d'Être of the magic numbers 28, 50, and 82. Accordingly, the variety of shape phase transitions including shape coexistence at and near the 8, 20, and 40 shell numbers could be explained within the framework of the spin-orbital interaction schemes [5-8]. We notify that the 70, one of the shell gap numbers of the harmonic oscillator potential, does not appear to be any shell gap character, because of a strong intruder $h_{11/2}$ orbital effect. Now we notice the emergence of deformed shape at N = 28. Though it is explained by the current shell-model schemes including, for instance, $2p_{3/2}$ and $1f_{7/2}$ orbitals between N = 28 shell gap, it is still uncertain what interactions force into combining the two orbitals, resulting in merging the $f_{5/2}$-$f_{7/2}$ spin-orbital doublet. If we follow that scenario, it is expected that a similar pattern should appear in the combined $d_{5/2}$ and $g_{9/2}$ orbital circumstances.

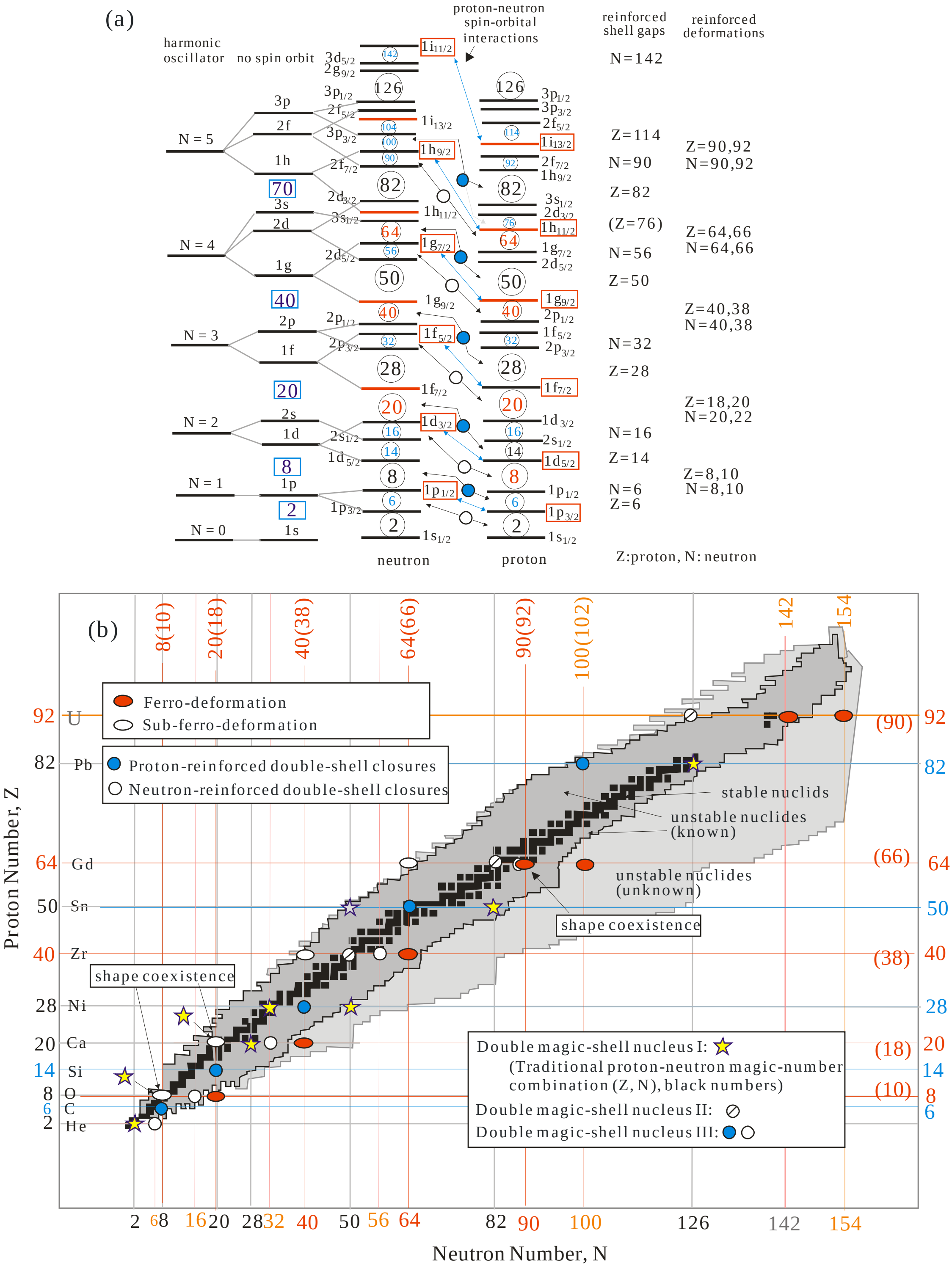

Fig. 2. (a) The evolution of shell model energy levels from the harmonic oscillator model with the quantum number N, through the single-particle energies of a Woods-Saxon potential, to the level splitting due to the spin-orbit coupling term. Note that the levels are numbered serially by a given orbital quantum number according to Nj = 2j + 1 of nucleons that can occupy each state. For discussions, we denote the associated spin-orbital doublet by the arrow and the corresponding critical configurations yielding spherical-shell closures. (b) Categorization of the nuclear shapes based on classification of the combination of the proton number and the neutron number.

## 3. Nuclear shapes panorama in the field 76 $\leq$ Z $\leq$ 84.

For more discussions regarding some emerging nuclear shapes, we investigate the systematic behavior of low-lying level properties in the region of 76 $\leq$ Z $\leq$ 84. Especially, the Z = 78, 80, and 82 nuclides are famous for emerging variety of shape coexistence in a nucleus; spherical, near spherical, non- spherical (deformed) shapes [9, 10]. Figure 3 illustrates the systematics of excitation energy ratio of the first $2^+$ and $4^+$ states, R = E($4^+$)/E($2^+$). This R value indicates a deformation parameter such that R < 2 for a spherical nucleus, R ~ 2 for a vibrator and R ~ 3.3 for a deformed axial rotor [12, 13].

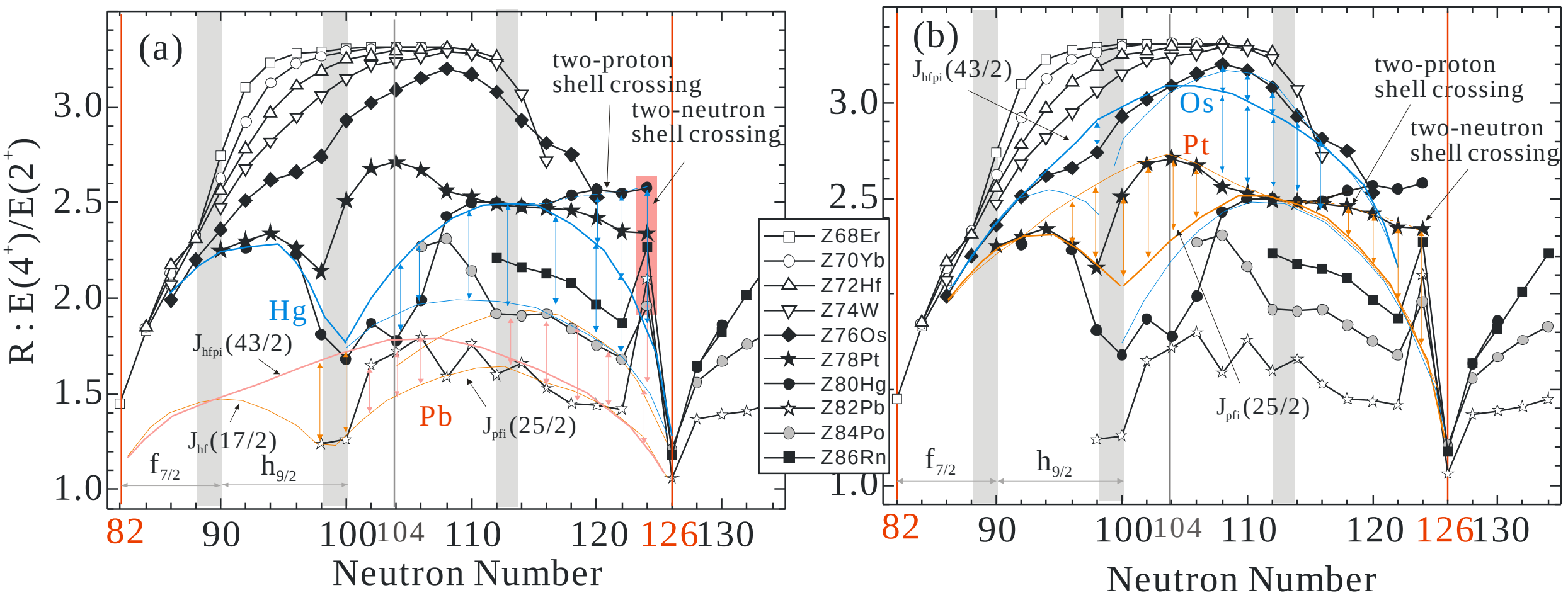

Fig. 3. Plots of nuclear deformation in terms of the deformation parameter R, the ratio of E($4^+$) to E($2^+$) for the Z = 68 to 86 space as a function of neutron numbers. Data are from the National Nuclear Data Center (NNDC) [11]. Lines in red and blue represent the assumed trajectories associated with the pseudo-shell shell configuration [5]. Vertical red and blue arrows indicate a possible shape coexistence for the nuclei at a given neutron number.

We firstly focus on the Z = 82 and 80; Pb, Hg. For the case of Z = 82, at N = 100 a shell gap is consistent with the expectation from the closure of pseudo-subshell $J_{hf}$(17/2). Though the formation of the pseudo-subshell $J_{hf}$(17/2), a broad peak, indicating another structure, appears to be around N = 104. This feature reflects the presence of one pseudo-shell centered at N = 104, $J_{hfpi}$(43/2). Thereby, it is revealed a complex pattern due to a competition between the one pseudo-shell and two sub-pseudo-shell configurations. Consequently shape coexistence would be expected within the space of N = 102 to 108, which are confirmed by observing the second $0^+$ states in the corresponding nuclei [9, 10]. For the case of the Z = 80, shapes are dramatically changed; at and around N = 100 a near spherical while at and above N = 108 a well deformed shape. More surprisingly, such a well deformed shape, with R ~ 2.5, continues to N = 124, at which should have occurred a near spherical shell-gap structure. Turning to the Z = 78 system, there is another dramatic change, which is a broad high-lying peak around N = 104. It is worthwhile again to emphasize that the number N = 104 corresponds to the center of a pseudo-shell $J_{hfpi}$(43/2). This characteristic, therefore, indicates that the formation of huge combined shell within N = 82 and 126 starts at Z = 78 and settles down at Z = 76. We see a remarkable contrast pattern showing a high peak and a deep valley, leading to a large hollow in between two protons, Z = 80 and 78. Taking into consideration of dynamical shell quantum effect, we expect a similar deformation at Z = 80 with that at Z = 78 and vice versa. *Here is the very presence of existing several shapes; a near-spherical or a non-spherical between two nuclei and the very shape coexistence in a nucleus.*

The reasons are very stereological: First, for the Z = 78, a strong deformation is centered at N = 104 since Z = 78 is a bit off the shell gap Z = 82, favoring a deformed configuration. The Z = 80 system, in contrast, yields a sub-pseudo-shell gap at N = 100 since it is readily affected by the Z = 82 shell gap. Accordingly, the Z = 80 favors two sub-pseudo-shell configurations, revealing the developed deformation around N = 114. Furthermore, the two (a pair of) protons crossing over the shell gap Z = 82 reinforces deformation with the two neutrons crossing over the N = 126. In other words, for the Z = 80, N = 124 system, both crossings of a pair of protons and neutrons force the surface toward a high-deformed shape, which is the reason of the occurrence of such a deformed peak at N = 124. For the case of Z = 78, we find that four shape coexistence appears; first is the $J_{hf}(17/2)$ phase, second is the $J_{pfi}(25/2)$ phase within N = 100 to 126, third is the $J_{hfpi}(43/2)$ whole phase, and fourth is at and below N = 124, a pair of neutron crossing phase. It is highly desirable that shape coexistence investigation should be done, for instance, for finding the second, third $0^+$ states. See the predicted shape coexistence as denoted the arrows in Fig. 3. At Z = 76, we observe a sudden change, indicating a phase transition toward the ferro-deformation regime with R ~ 3.3. Nonetheless, a shell gap effect due to N = 100 is still present so that shape coexistence would be expected in a region of N = 96 to N = 104. At Z = 74, we finally arrive on the ferro-deformation regime [5].

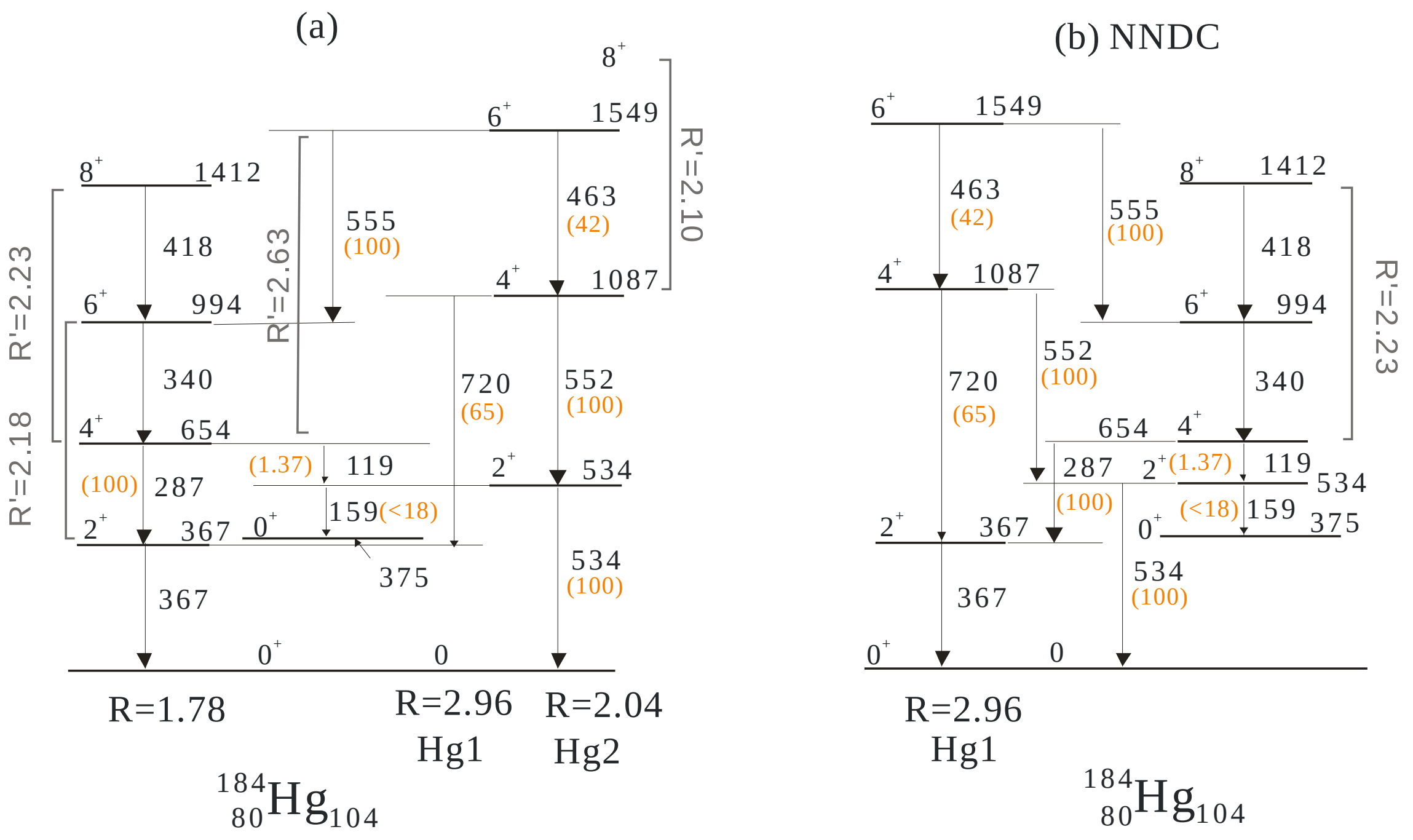

Fig. 4. Partial level schemes of $^{184}$Hg. Data are taken from the national nuclear dater center (NNDC) [11]. (a) The scheme in the present work. (b) The scheme based on NNDC. For a comparison between the R values concerned, we denote Hg1 and Hg2, which are from the ratios of the $4^+$ at 1087 keV to the $2^+$ at 367 keV and to the $2^+$ at 534 keV, respectively. Notice that the quoted R values correspond to the energy ratios of the *first* excited $4^+$ and $2^+$ states that is, in the case of $^{184}$Hg, R = 1.78. We add the R-like values, R', based on the higher lying levels such as the ratio of $6^+$ to $4^+$ states, which provide more evidence for the corresponding band structure. We can see two bands representing a vibrational structure as indicating strongly mixed features with near spherical and non-spherical states between the ground state and the second $0^+$ state. The numbers in parentheses are the branching ratios of gamma rays depopulating the corresponding levels [11]. Fruitful discussions, including many references, on the shape coexistence are found in [9, 10].

We again study shape coexistence for more details as focusing on $^{184}$Hg, with N = 104. Figure 4 illustrates the partial level schemes of $^{184}$Hg based on the NNDC data base. According to the level scheme provided by NNDC, as shown in Fig. 4(b), the structure would be classified into two; one is a large deformation with R = 2.96 built on the ground state, as denoted Hg1, and the other is built on the second $0^+$ state. In contrast, by reshaping the given level scheme, as shown in Fig. 4(a), we offer an evidence for showing another structure built on the ground state indicating R ~ 2 as well as R = 1.78. As already pointed out, the quoted R values are based on the first $4^+$ and $2^+$ states in the corresponding nuclides, namely R = 1.78 for the case of $^{184}$Hg. Most of discussions regarding shape coexistence for the Hg nuclides in literature were made with the Fig. 4(b) level scheme base. Thereby the structure based on our R value parameters would not be revealed but *concealed* [9, 10]. We now redraw the R value systematics in order to make sure of our concept in Fig. 5, where another region systematics is added for a comparison.

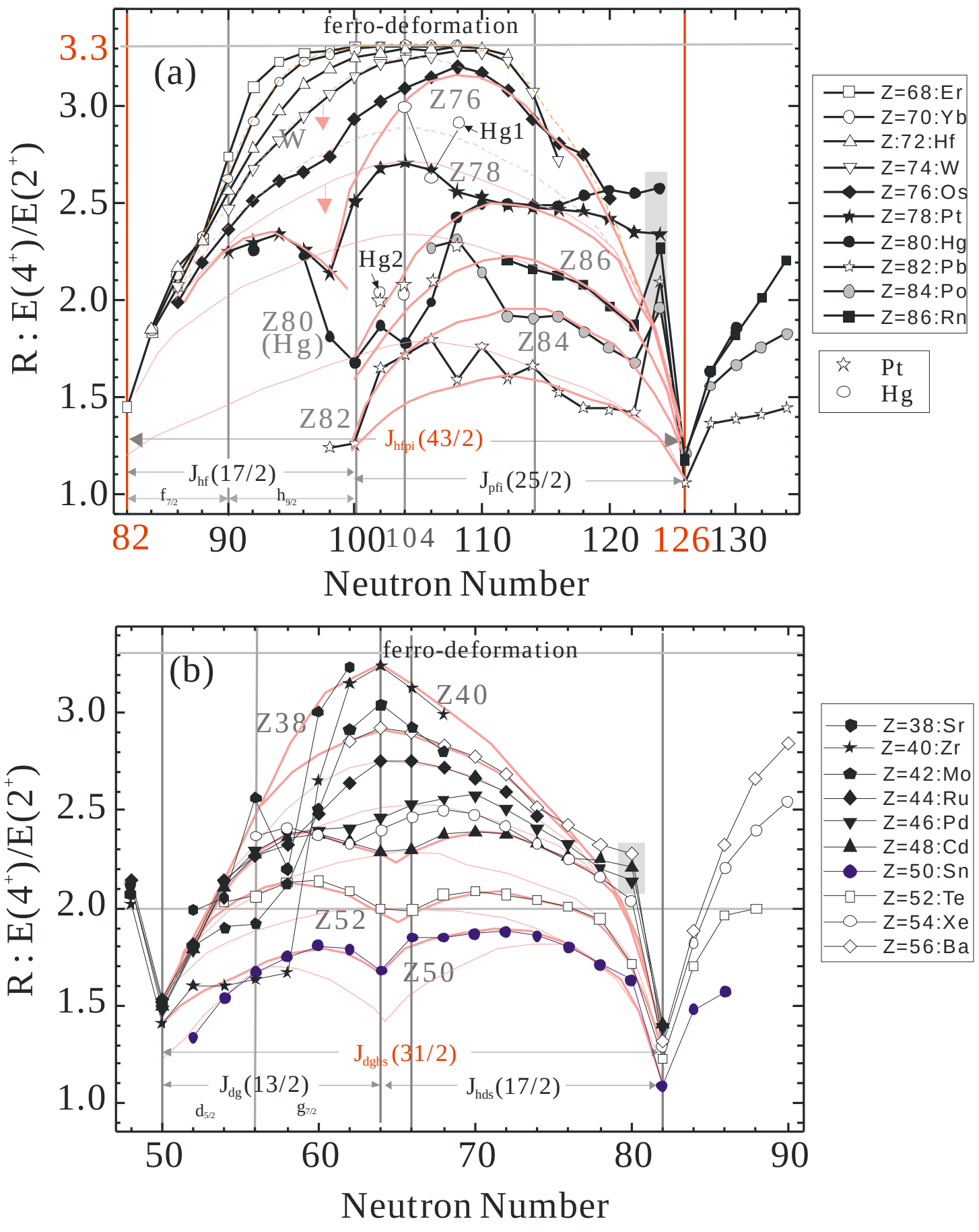

Fig. 5. Plots of nuclear deformation R values, for (a) the Z = 68 to 86 space and (b) the Z = 38 to 56 space as a function of neutron numbers. Data are from the national nuclear data center [11]. Lines in red represent the associated trajectories with the pseudo-shell formation. The $J_{hf}(17/2)$, as an example, represents the pseudo-subshell combined with the two $f_{7/2}$ and $h_{9/2}$ orbitals, having J = 17/2 capacity. See for more details Ref. [5]. In (a), the open circles for Hg represent the R values obtained as shown in Fig. 4 while the open stars for Pt do the R values associated with the band structure built on the second $2^+$ state at 102, 104, 106, and 108, respectively. We can see the close similarities and clear differences between two regimes in (a) and (b). It is worthwhile to understand, as noticing to the assumed trajectories denoted red (dotted) lines in (a), that there are dynamical shape crossings and transitions in between isotopes, isotones, and isobars. Our model based on the pseudo-shell configuration plays a central role in predicting and reshaping nuclear shape coexistence.

We agree to the provided results [9, 10] that the second $0^+$ state is a band-head of a deformed structure, indicating R ~ 2.2 as shown in Fig. 4. And also it is well understood that there are some strong mixed states between the ground level and the excited levels at the second $0^+$, the first $2^+$, $4^+$ and second $2^+$, $4^+$ states. Nevertheless, a stabilized structure emerged in both two bands indicates a near-spheroidal shape associated with a vibrator rather than a rotator; one with R ~ 2.0 and the other with R ~ 2.2. As can be seen in Fig. 5(a), the experimental data, based on Hg2 scheme, are in agreement with our expected configuration as obtained by the systematic trends. Moreover, the Hg2 values of Z = 78, Pt, are close to the values of Hg2 of Hg, which provides us with a remarkable shape coexistence between the Z = 80 and 78 systems around N = 104. The Hg1 structure is also obtained at N = 106 and 108 if we follow the level schemes provided by NNDC, which lies in R ~ 2.8. However, such a large deformation is less likely since the R = 2.8 is too high to be fitted with the systematic line for Z = 80. Our conclusions are as follows; the Z = 80 system, Hg, represents a near spherical vibrator at N = 102, 104, and 106. Around N = 114, the center of the pseudo-subshell $J_{pfi}(25/2)$, it shows an intermediate deformed structure, indicating rather a rotator. At and near N = 124, it has a different deformation derived by the two-neutron crossing over N = 126 shell closure. Finally it reveals a mixed deformed structure within the space N = 114 and 124, affected by both the two-proton and the two-neutron crossings, respectively, over Z = 82 and 126 shell closures. We summarize our results in Fig. 5, where a variety of shape phases and shape transitions are exhibited for the space Z = 68 to 86 and for the space Z = 38 to 56. It is interesting to compare the similarity and the difference of level structures between the Z = 80, 84 (Hg, Po) systems and the Z = 48, 52 (Cd, Te) systems.

## 3. Conclusions.

We describe schematically the nuclear shape phases over the nuclear chart along with shell model schemes. Various nuclear shapes are demonstrated from the results of proton-neutron spin-orbital interactions; ferro-deformed, sub-ferro-deformed, and spherical shapes. The spherical shape is classified into the three categories in view of large shell gap schemes; double-magic nuclei I, double magic nuclei II, and double magic nuclei III. We also demonstrate a variety of nuclear shapes in the space Z = 76 to 84. Fruitful and dynamical shape phases are introduced, by putting a comment on the nuclear shape transitions and shape coexistence.